\documentclass{llncs}

\usepackage{color}
\usepackage{amsmath}
\usepackage{multirow}
\usepackage{booktabs}
\usepackage{url}
\usepackage{amsmath}
\usepackage{subfig}
\usepackage{adjustbox}
\usepackage{subfloat}
\usepackage{xcolor}
\usepackage{mdframed}
\usepackage{framed}
\usepackage{tabularx}
\usepackage{comment}

\usepackage{lipsum}
\usepackage{graphicx}

\usepackage{float}
\restylefloat{table}

\DeclareUnicodeCharacter{00B4}{'}

\newcommand\blfootnote[1]{%
  \begingroup
  \renewcommand\thefootnote{}\footnote{#1}%
  \addtocounter{footnote}{-1}%
  \endgroup
}

\begin{document}

\title{Designing User-Centered Simulations of Leadership Situations for Cave Automatic Virtual Environments: Development and Usability Study}

\author{Francesco Vona\inst{1}\orcidID{0000-0003-4558-4989} \and Miladin Ćeranić\inst{1}\orcidID{0009-0005-4700-1240} \and Irma Rybnikova\inst{1}\orcidID{0000-0002-1162-755X} \and Jan-Niklas Voigt-Antons\inst{1}\orcidID{0000-0002-2786-9262}}

\institute{Immersive Reality Lab, University of Applied Sciences Hamm-Lippstadt, Hamm, Germany \\ \email{jan-niklas.voigt-antons@hshl.de}}

\maketitle
\blfootnote{This paper has been accepted for publication in the Human-Computer Interaction International conference 2023. The final authenticated version is available online at https://doi.org/10.1007/978-3-031-36004-6\_45.}
\thispagestyle{empty}
\pagestyle{empty}

\begin{abstract}
Given that experience is a pivotal dimension of learning processes in the field of leadership, the ongoing and unresolved issue is how such experiential moments could be provided when developing leadership skills and competencies. Role-plays and business simulations are widely used in this context as they are said to teach relevant social leadership skills, like those required by everyday communication to followers, by decision-making on compensation, evaluating performance, dealing with conflicts, or terminating contracts. However, the effectiveness of simulations can highly vary depending on the counterpart's ability to act in the given scenarios. In our project, we deal with how immersive media could create experiential learning based on simulations for leadership development. In recent years different variations of extended reality got significant technological improvements. Head-mounted displays are an easy and cost-efficient way to present high-resolution virtual environments. For groups of people that are part of an immersive experience, cave automatic virtual environments offer an excellent trade-off between actual exchange with other humans and interaction with virtual content simultaneously. The work presented is based on developing a user-centered simulation of leadership situations for cave automatic virtual environments and includes the results of a first usability study. In the future, the presented results can help to support the development and evaluation of simulated situations for cave automatic virtual environments with an emphasis on leadership-related scenarios.


\end{abstract}

\section{Introduction}
The development of effective leadership skills and competencies has long been recognized as a crucial aspect of organizational success and individual growth \cite{northouse2018leadership}. However, the process of cultivating these capabilities often hinges on experiential learning, which can be challenging to provide in a controlled and targeted manner \cite{carroll2022experiential}. Traditional methods such as role-plays and business simulations have been widely employed to teach essential social leadership skills \cite{kolb2014experiential}, but their efficacy can be significantly impacted by the performance and adaptability of the individuals involved. As a result, there is an ongoing need to explore novel approaches for delivering experiential learning opportunities in leadership development.

Despite repeated calls and appeals to consider immersive media as a new and potentially fruitful approach to leadership development (e.g. \cite{guthrie2011developing}), there is a lack of empirical investigations that aim at developing, testing and evaluating concrete learning tools based on immersive media with the focus of leadership development. With our study, we address this shortcoming.

The rapid advancements in extended reality (XR) technologies, including head-mounted displays and cave automatic virtual environments (CAVEs), offer promising new avenues for enhancing the learning experiences of aspiring leaders \cite{slater2009place}. The immersive nature of these technologies enables users to engage in realistic and interactive simulations, providing valuable opportunities for practicing leadership skills in a controlled environment \cite{blascovich2011infinite}. This paper focuses on the potential application of CAVEs in creating user-centered simulations of leadership situations, addressing the existing gap in experiential learning opportunities.

The motivation behind our research stems from the recognition that immersive media could revolutionize the way leadership skills are developed and practiced. By leveraging the unique capabilities of CAVEs, we aim to provide an effective, interactive, and engaging platform for leadership development, enhancing the overall learning experience \cite{guthrie2011developing}. Our work involves the design and implementation of a user-centered simulation tailored for CAVE environments, as well as an initial usability study to assess the effectiveness of our approach.

Ultimately, the findings presented in this paper hold the potential to significantly impact the future of leadership development, providing valuable insights and guidance for the creation and evaluation of simulated scenarios within CAVE environments \cite{bowman2007virtual}. By harnessing the power of immersive technologies, we hope to contribute to a more effective, engaging, and accessible approach to leadership education and training, addressing the current lack of empirical investigations in the field \cite{guthrie2011developing}.

\section{Related Work}
This section provides an overview of the research landscape relevant to immersive media, tracing its roots from the initial theories of flow to the development of CAVE systems for media presentation.

\textbf{Flow Theory:} The concept of flow, first introduced by \cite{csikszentmihalyi1975}, refers to a mental state of complete immersion and absorption in an activity. In this state, individuals experience heightened focus, deep involvement, and optimal enjoyment. In the context of media experiences, flow is considered an essential component for achieving immersion and engaging users effectively \cite{ghani1994}.

\textbf{Immersion and Presence in Virtual Environments:} The concept of presence, or the subjective feeling of being in a virtual environment, has been extensively studied in relation to immersive media. \cite{steuer1992} identified three factors that contribute to the sense of presence in virtual environments: vividness, interactivity, and the extent to which the user can control the environment. Lombard et al.\cite{lombard1997} proposed six dimensions of presence, including spatial presence, social presence, and self-presence, which have since been used to inform the design of immersive media experiences. High immersive systems can be used to evoke strong emotions \cite{voigt2021don}. Bio-signals that are related to emotions can be measured in real time and used to adapt content of immersive media systems \cite{pinilla2023realtime}. Besides technological factors, also social environments can have an influence on user experience and social acceptability of immersive systems \cite{vergari2021influence}.



\textbf{CAVE Systems for Media Presentation:} Cave Automatic Virtual Environment (CAVE) systems, introduced by \cite{cruz-neira1992}, represent a significant advancement in the immersive media presentation. CAVE systems consist of large-scale, room-sized displays that surround users with multiple projection screens, creating an enveloping 3D visual experience \cite{cruz-neira1992}. These systems have been widely used in scientific visualization, architectural design, and cultural heritage applications \cite{carrozzino2010}. The inherent advantages of CAVE systems over HMD-based VR, such as the allowance for natural group collaboration and the elimination of motion sickness, have led to their continued relevance and application in various research and industrial contexts \cite{shaw2014}.

In conclusion, the study of immersive media has evolved from foundational theories of flow and presence to the development of advanced technologies such as VR and CAVE systems. The continued exploration of these technologies will undoubtedly result in novel applications and further insights into the potential of immersive media experiences.

In the field of leadership development, experience, and experience-based tools are considered as the most effective approach to obtaining leadership skills and competencies. In conceptual terms, this idea goes back to the experiential learning theory by \cite{kolb1984experiential}. The author states that learning processes could be described as processes of meaning-making from personal experiences. Especially those experiences that are assumed to trigger meaning-making and learning that challenge, irritate, or unsettling. One of the established ways to provide such experiential learning when developing leaders are business games and role-plays. They are widely used in leadership development settings. Leadership students are said to more effectively achieve relevant leadership skills, like those required by everyday interactions with followers, decision-making and decision communicating to followers, or conflict resolution, than using case studies \cite{dentico1999leadsimm}. The main reason is a high level of realism and high involvement of students in business games \cite{hunsaker1977leadership}. 
Although immersive media is a flourishing field and bear a particularly high potential for experiencing learning, there is barely research on how immersive media could be used in leadership development. Instead, there are several studies that investigate online gaming as a potentially educative tool to develop effective leaders. Reeves et al.\cite{reeves2008leadership} argue that multiplayer-online games can teach leadership competencies that are needed in business companies. Some scholars appeal to consider game-based virtual worlds as a “prospective avenue for leadership skill development” since they require a wide array of skills and competencies that are of great value for leadership tasks, like role delegation, member motivation, persuasion, and collaborative work \cite{guthrie2011virtual}. Virtual reality is also said to hold manifold potential for leadership development since intelligent tutoring tools may be able to adapt situations to the learner’s behaviors. Studies argue in favor of the effectiveness of educational tools based on virtual reality and point out that especially avatar-based games provide a psychologically secure environment to test and improve the interactive skills of students in an almost-real way \cite{fecke2022simulationen}. 

\section{Methods}
\textbf{Participants:} Ten people participated in the study (8 self-identified as male and two self-identified as female). The mean age was 30.1 years, with a standard deviation of 5.8. The experiment was approved by the local ethics committee.

\textbf{Procedure:} Participants were invited to a lab room at a different separated time slot to participate alone in the experiment.
In the beginning, the participants were welcomed by a moderator and presented with an introduction to the study and its purpose. After signing a consent form, the participants were given a pre-questionnaire asking about demographic information and their affinity for technology interaction \cite{franke2019personal}.  
The next step was to introduce participants to the simulated leadership situations. After this introductory part, the participant could start the condition communicated by the moderator. The participants were supposed to play each of the four conditions in randomized order. After each condition, the participant was asked to answer web-based questionnaires encompassing User Experience Questionnaire (UEQ), igroup Presence Questionnaire (IPQ), and Social Presence Questionnaire (SPQ) \cite{Schubert_undated-to,Laugwitz2008-ab,socialpresence}. After all the conditions had been played and rated, the participant was asked to rate the overall experience and which condition they liked most. The complete duration of the experiment was around 40 minutes.

\textbf{Apparatus:} A CAVE (CAVE Automatic Virtual Environment) is a room where the environment is simulated in three dimensions for an observer. The walls in such a room are often made of projection screens on which content can be displayed. 
CAVE projection systems can consist of up to six projection screens for simulations in a cuboid-shaped room. The CAVE used at the Hamm-Lippstadt University of Applied Science consists of four acrylic glass screens: a front projection, two side projections, and a floor projection screen. 
Four 3D special projectors are used to project the image, which is redirected onto the projection screens via mirrors due to space limitations in the room. To track the observer's movements, a marker tracking system consists of fixed and mobile infrared cameras. A sound system consisting of subwoofers and passive speakers is also installed.
\textbf{Virtual Reality Application:} The Virtual Reality application was created in Unity and consists of 4 scenarios, and 2 simulated leadership situations, each with animated and non-animated virtual avatars (2x2). All scenarios are composed of at least two people (one real), and the user must impersonate a different character each time (once the boss, and once the employee, for each situation). The 2 simulated leadership situations concern the themes of "giving critical feedback as a supervisor" and "leadership and employee health".
In the first scenario, there are 2 people: Tim and Mike. Tim is responsible for the payroll department of human resources. After the introduction of new software that allows for more efficient payroll accounting, Tim is not satisfied with Mike's performance because he is not yet proficient in using the software. 
Now, in the Human Resources department, the annual staff evaluation is scheduled. In this scenario the user has to play once Tim and the other Mike.

\begin{figure}[!htb]
\includegraphics[width=0.5\textwidth]{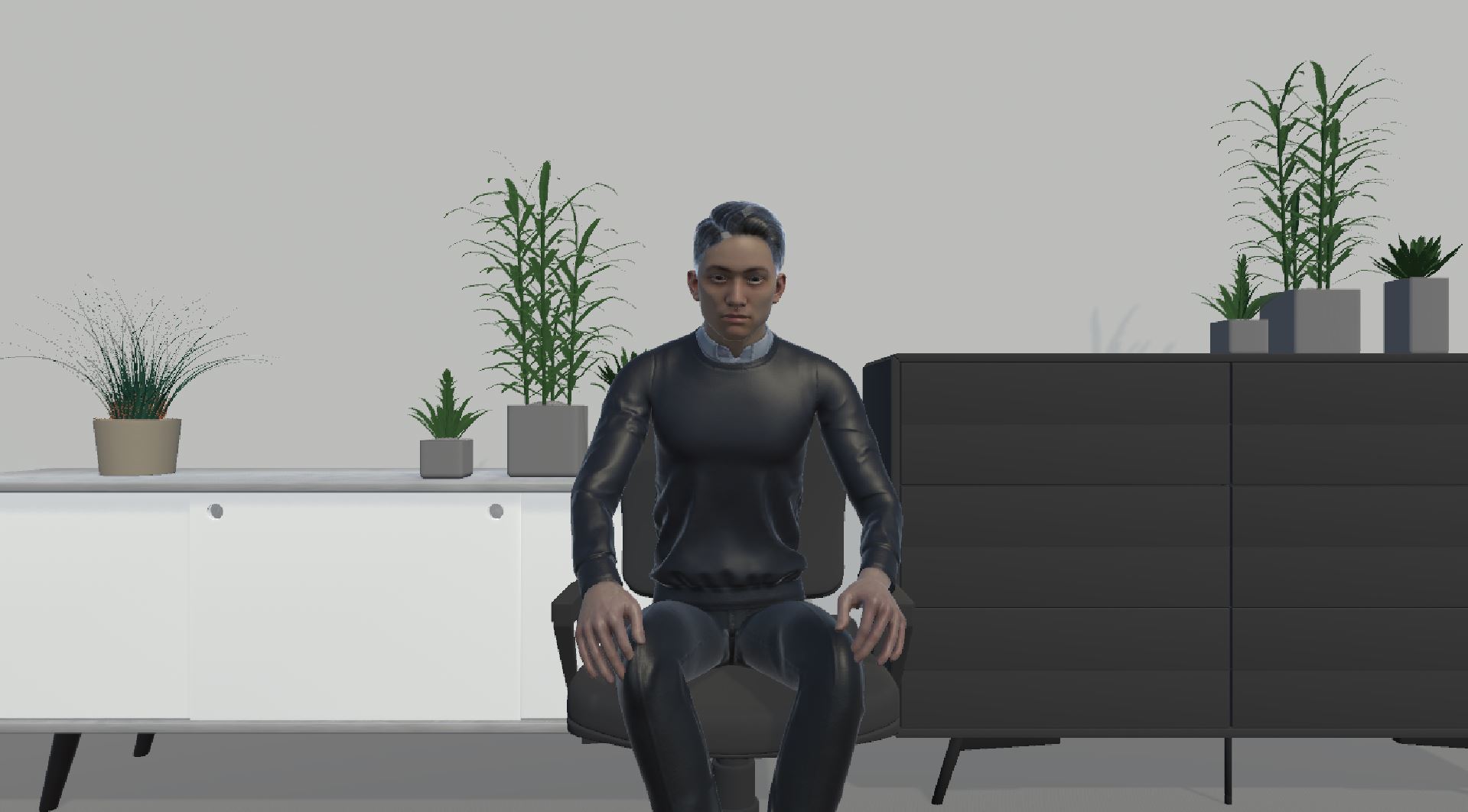}
\includegraphics[width=0.5\textwidth]{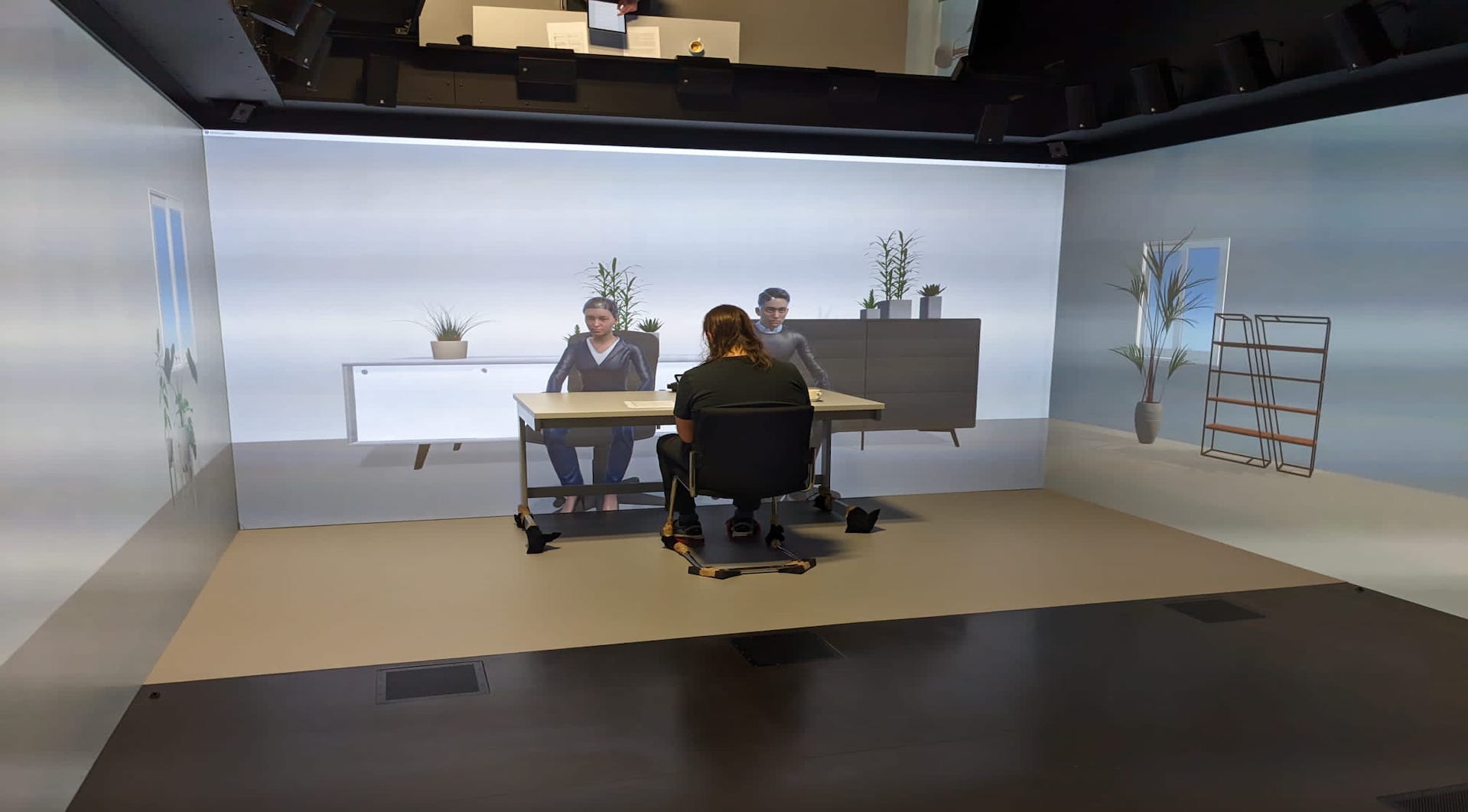}
\caption{On the left: what the user sees in the virtual environment; on the right: a picture taken from outside the CAVE }
\label{fig:scene}
\end{figure}

The second scenario concerns workplace health. In German organizations, there are mental risk assessments for every workplace. This is not a measurement of employees' mental health but an analysis to determine whether work activities, the workplace, or the environment in which it takes place are placing employees under stress or could lead to psychological stress or illness. The idea behind this evaluation is to prevent illnesses and inform employees and employers on how to design working conditions so that employees feel well. 
In this scenario, there are three characters: a manager (Tom) and two employees (Anne and Sören).
In this scenario, the user must interpret Tom and Sören once each.
All four scenarios take place in the same virtual environment: a very simple office with some furnishings in the background (furniture and plants). Figure \ref{fig:scene} shows the experiment with a first person and third person point of view. All avatars have audio that can be played from the moderator's keyboard during the experience. In this way, the experiment moderator can control the entire execution of the experience. Virtual characters will have animations during audio playback in conditions with animated avatars, while in conditions with non-animated avatars, avatars remain in a sitting idle position. Avatars and animations were taken from Mixamo \cite{mixamo}.

\section{Results}
\subsection{UEQ-S}
\begin{figure}[!htb]
\includegraphics[width=0.5\textwidth]{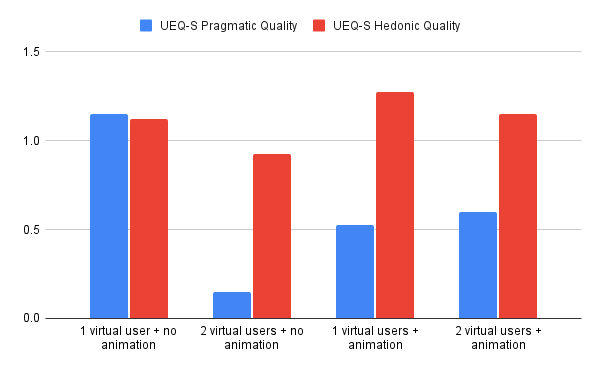}
\includegraphics[width=0.5\textwidth]{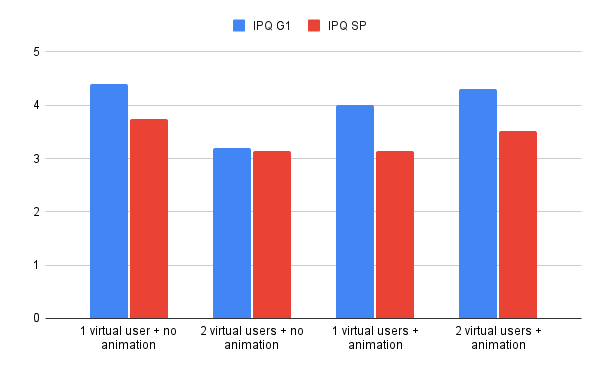}
\caption{On the left: Average values over all participants for the two UEQ-S scales pragmatic and hedonic quality. On the right: Average values over all participants for the two IPQ scales general presence (IPQ G1) and spatial presence (IPQ SP).}
\label{fig:ueq-ipq}
\end{figure}

The results for the UEQ-S dimensions of the pragmatic and hedonic quality can be found in Figure \ref{fig:ueq-ipq}. For condition 1 (1 virtual user + no animation) hedonic quality was similar to the pragmatic quality (PQ = 1.15, HQ = 1.12). For condition 2 (2 virtual users + no animation) pragmatic quality was rated as being lower compared to the hedonic quality (PQ = 0.15, HQ = 0.92). A similar trend could be observed for condition 3 (1 virtual users + animation) and condition 4 (2 virtual users + animation), for which also pragmatic quality was rated as being lower compared to the hedonic quality (condition 3: PQ = 0.52, HQ = 1.27; condition 4: PQ = 0.6, HQ = 1.15).

\subsection{IPQ}
For the one virtual user condition (condition 1), the perceived overall presence (G1) was reduced when adding animations from a mean value of 4.4 to 4. Adding animation for the one user condition resulted in a lower experienced spatial presence, from 3.74 to 3.14. In the case of the two-user conditions, adding animation resulted in a higher perceived overall presence (from a mean value of 3.2 to 4.3) and also a higher spatial presence (from 3.14 to 3.42).

\subsection{After condition questions}
A 7-value Likert scale was used for the evaluation of the post experiment questions.
The results of the questions also show that the first condition, with an avatar and no animation was the one felt most real with a mean value of 4,3. This is probably due to the fact that the avatar in the idle position seemed to maintain eye contact with the user which in the same condition with animation and with two users did not happen. For the second question, in addition to condition 1 (mean = 5.7),also conditions 2 (mean = 5.5) and 4 (mean = 5.6) scored high. This may have to do with the role played during the scenarios.
\section{Conclusion}
To summarize, it is crucial to exercise prudence in situations involving a single virtual user. Incorporating non-congruent animations may negatively affect both overall and spatial presence perception. As such, it is essential to thoughtfully choose animations and details. In contrast, the presence of two virtual users seems to enhance the experience, as the addition of animation appears to have a generally positive influence on presence.

The vast potential of immersive media for experiential learning and training in leadership cannot be understated. Nonetheless, educational institutions must recognize the significant time and effort necessary for the development, testing, and evaluation of such tools. Furthermore, forging a more robust connection between leadership theories and innovative leadership development tools presents an opportunity for future research and practical applications.   

\bibliographystyle{splncs04}
\bibliography{references.bib}

\end{document}